\documentstyle[12pt]{article}
\catcode`\@=11
\def\@mystartsection#1#2#3#4#5#6{\if@noskipsec \leavevmode \fi
   \par \@tempskipa #4\relax
   \@afterindenttrue
   \ifdim \@tempskipa <\z@ \@tempskipa -\@tempskipa \@afterindenttrue\fi
   \if@nobreak \everypar{}\else
     \addpenalty{\@secpenalty}\addvspace{\@tempskipa}\fi \@ifstar
     {\@ssect{#3}{#4}{#5}{#6}}{\@dblarg{\@sect{#1}{#2}{#3}{#4}{#5}{#6}}}}
\def\section{\@mystartsection{section}{1}{\z@}
{-3.5ex plus -1ex minus -.2ex}{2.3ex plus .2ex}{\Large\bf}}
\def\subsection{\@mystartsection{subsection}{2}{\z@}
{-3.25ex plus -1ex minus -.2ex}{1.5ex plus .2ex}{\large\bf}}
\def\subsubsection{\@mystartsection{subsubsection}{3}{\z@}
{-3.25ex plus -1ex minus -.2ex}{1.5ex plus .2ex}{\normalsize\bf}}
\def\be{\begin{equation}}
\def\ee{\end{equation}}
\def\beq{\begin{eqnarray}}
\def\eeq{\end{eqnarray}}
\def\nin{\noindent}
\newcounter{refs}
\begin{document}
\def\rref#1{(\ref{#1})}

\flushbottom
\pagestyle{empty}
\setcounter{page}{0}
\rightline{ DFTT 67/92}
\rightline{ November 1992}
\rightline{ hep-ph/9212247}
\vspace*{1cm}
\centerline{\LARGE\bf Neutrino mixing}
\vspace*{1cm}
\centerline{\Large S.M. Bilenky }
\vspace{0.5cm}
\centerline{\large\it
Joint Institute of Nuclear Research, Dubna, Russia }
\centerline{\large\it and }
\centerline{\large\it
Istituto Nazionale di Fisica Nucleare, Sezione di Torino }
\centerline{\large\it
Via P. Giuria 1, I--10125 Torino, Italy }
\vspace*{0.5in}
\centerline{\Large Abstract }
\begin{quotation}
In the first part of these lectures the neutrino mixing hypothesis
will be considered in detail. We will discuss the possible schemes
of neutrino mixing and present the data of the recent experiments
searching for effects due to nonvanishing neutrino masses and mixing
angles.In the second part of these lectures the physics of solar
neutrinos will be considered.We will discuss the MSW resonance
solution of the equation of evolution of a neutrino in matter and
present data of solar neutrino experiments.
\end{quotation}
\vfill
\newpage
\pagestyle{plain}

\section{Introduction}

\indent
One of the most important question in neutrino physics is the problem
of neutrino masses and mixing.From the point of view of many models
beyond the standard one it is very natural for neutrinos to be massive.
In the minimal standard model with only left-handed neutrino fields,
neutrinos are massless particles.However, the standard model can be
easily generalized if we assume that singlet right-handed neutrino
fields are presented in the Lagrangian.In this enlarged standard model
neutrinos,like leptons and quarks, are massive particles.

Leptons and quarks are electrically charged Dirac particles (particle
$\neq$ antiparticle).For massive neutrinos there are
 two possibilities.
Neutrinos with definite masses can possess  some conserved
lepton number and be Dirac
particles or, in theories where there is no conserved lepton numbers,
massive neutrinos are truly neutral Majorana particles(particle
$\equiv$ antiparticle).
Theories with massive Majorana neutrinos are beyond the standard theory.

If neutrinos are particles with Dirac or Majorana masses their fields
can appear in the weak currents in mixed form.This is the so called
neutrino mixing hypothesis $[{\ref{ref.1},\ref{ref.2}}]$.
Mixing of fermion fields is a characteristic feature of modern gauge
theories with spontaneous violation of symmetry. The Cabibbo-
Kobayashi- Maskawa mixing of quarks is a well known phenomenon.
Does neutrino mixing take place too? In more than 50 experiments
the problem of neutrino mixing is being investigeted by different
methods.Up to now no indications in favour of nonzero neutrino
masses and mixing was obtained in experiments with terrestrial
neutrinos.Experiments are continuing, however, and in the nearest
future new level of accuracy will be achived in experiments
searching for $\nu_\mu\ \rightarrow \nu_\tau$
oscillations, neutrinoless double $\beta$-decay and other
experiments.

The solar neutrino experiments play a special role in the test of
the neutrino mixing hypothesis.These experiments are sensitive to
values of neutrino masses and neutrino mixing angles so small
that could not be reached in experiments with reactor and
accelerator neutrinos.

In the first part of these lectures we will consider possible
schemes of neutrino mixing and possible experiments to search
for effects of neutrino masses and mixing.We will present also
data of some of the most recent experiments.In the second part
we will discuss the solar neutrino experiments.Resonance transitions
of neutrinos in matter (MSW-mechanism) $[{\ref{ref.3}}]$ will be
considered in detail. As an introduction, it seems appropriate to
recall the standard Higgs mechanism for the generation of quark masses.

\section{Higgs mechanism of quark mass generation.}

Let us introduce the doublet of scalar Higgs fields

\begin{equation}
\phi={\phi_{+}\choose\phi_{0}}
\label{eq:1}
\end{equation}

\noindent and assume that the $SU(2) \times \nonumber U(1)$ invariant
Lagrangian of interaction of quarks and Higgs boson has
the Yukawa form

\begin{equation}
{\cal{L}}=-\frac{\sqrt 2}{v}
\sum_{\stackrel{i=1,2,3}{q=d,s,b}}
\bar \psi_{iL} M_{iq} q'_{R} \phi~~+~~h.c.
\label{eq:2}
\end{equation}

\noindent Here $\psi_{1L}={u'_{L}\choose d'_{L}},
{}~\psi_{2L}={c'_{L}\choose s'_{L}},
{}~\psi_{3L}={t'_{L}\choose b'_{L}}$ are doublets of quark fields, $q'_{R}$
are right-handed singlets ,$M $ is a complex nondiagonal $ 3 \times 3 $
matrix,$v $ is a parameter.After spontaneous violation of symmetry we can
put (unitary gauge)

\begin{equation}
\phi(x)={0\choose{\frac{v+\chi(x)}{\sqrt 2}}},
\label{eq:3}
\end{equation}

\noindent where $ \chi(x) $ is the field of the scalar, neutral Higgs
particles.An arbitrary complex matrix M can be diagonalized with the
help of a biunitary transmormation:

\begin{equation}
M=V_{L}mV_{R}^{+},
\label{eq:4}
\end{equation}

\noindent where $V_{L} $ and $V_{R} $ are $3 \times 3 $ unitary matrixes;
$m_{ik}=m_i\delta_{ik} $, $m_i>0 $. From Eq.(2), Eq.(3)
and Eq.(4) we have

\begin{equation}
{\cal{L}}=-\sum_{q=d,s,b} m_q \bar{q} q-\sum(\frac{m_q}{v})\bar{q} q\chi,
\label{eq:5}
\end{equation}

\noindent where

\begin{equation}
\left( \begin {array}{c}
d'_{L,R}\\s'_{L,R}\\b'_{L,R}
\end{array} \right)
=V_{L,R}
\left( \begin{array}{c}
d_{L,R}\\s_{L,R}\\b_{L,R}
\end{array} \right)
\label{eq:6}
\end{equation}

\noindent The first term in expression Eq.(5) is the mass term of "down"
quarks.As it is seen from Eq.(6), the L(R) components of the fields that
enter into the initial multiplets are connected with the L(R)
components of quarks fields with definite masses by unitary
transformations. Analogously, for the fields of "up" quark we have

\begin{equation}
\left( \begin {array}{c}
u'_{L,R}\\c'_{L,R}\\t'_{L,R}
\end{array} \right)
=U_{L,R}
\left( \begin{array}{c}
u_{L,R}\\c_{L,R}\\t_{L,R}
\end{array} \right),
\label{eq:7}
\end{equation}

\noindent where $U_{L,R} $ are unitary matrixes.

Further, the standard charged current is given by

\begin{equation}
j^W_{\alpha }=2[\bar{u}'_{L}\gamma_{\alpha}d'_{L}+
\bar{c }'_{L}\gamma_{\alpha}s'_{L}+\bar{t}'_{L}\gamma_{\alpha}b'_{L}]
\label{eq:8}
\end{equation}

\noindent Using Eq.(6) and Eq.(7) we can rewrite current
$j_{\alpha}^W $ in terms of physical quark fields as follows

\begin{equation}
j_{\alpha}^W=2[\bar{u}_{L}\gamma_{\alpha}d_L^c +
\bar{c}_L\gamma_{\alpha}s_L^c + \bar{t}_L\gamma_{\alpha}b_L^c]
\label{eq:9}
\end{equation}

\noindent Here

\begin{equation}
d_L^c=\sum_{q=d,s,b }V_{uq}q_L~,~s_L^c=\sum_{q=d,s,b}V_{cq}q_L~,~
b_L^c=\sum_{q=d,s,b }V_{tq}q_L
\label{eq:10}
\end{equation}

\noindent and $V=V_L^{+}U_L $ is the unitary mixing matrix.

Thus, in the general case of spontaneous violation of symmetry
quark fields appear in the charged current in mixed
form.\footnote{Let us notice that due to the unitarity of the matrices
$V_{L,R} $ and $U_{L,R} $ the standard neutral current is diagonal in
the quark fields.}

Now let us turn to the subject our lectures,namely the lepton sector.
The standard lepton charged current has the form

\begin{equation}
j_{\alpha}=2\sum_{l=e,\mu,t}\bar{\nu}_{lL}\gamma_{\alpha}l_L
\label{eq:11}
\end{equation}

\noindent What are neutrino fields $\nu_{lL} $ ? Are they real flavour
neutrino fields or mixture of fields of neutrinos with definite masses:

\begin{equation}
\nu_{lL}=\sum_{i}U_{li}\nu_{iL},
\label{eq:12}
\end{equation}

\noindent ($\nu_i $ is the field of neutrinos with mass $m_i $, U is
 unitary matrix). This last assumption is the so called neutrino mixing
hypothesis. In the next chapter we will consider different possibilities
of neutrino mixing.

\section{Schemes of neutrino mixing.}

\subsection{Dirac mass term.}

\indent
Schemes of neutrino mixing are usually  characterized by the type of the
relevant mass terms.From Lorentz invariance it follows that in general
three possible neutrino mass terms can be built $[{\ref{ref.4}}]$. All
of them may appear in different gauge models. In this section we will
consider the mixing scheme, that corresponds to Dirac mass term.

\begin{equation}
{{\cal{L}}^D}=-\sum_{l',l=e,\mu,\tau }\bar{\nu}_{l'R}
 M_{l'l}\nu_{lL}~~+~~h.c.,
\label{eq:13 }
\end{equation}

\noindent that is analogous to the quark mass term (see Eq.(5))
and could be generated by the standard Higgs mechanism we discussed
in the previous paragraph. In the expression Eq.(13)
M is a 3 $ \times$ 3  complex, nondiagonal matrix,
$\nu_{lL }$  are neutrino
fields that appear in the standard charged and neutral currents
(current fields). We will assume that right-handed fields $ \nu_{lR}$
enter only into mass term ${{\cal{L}}^D} $. After the standard
procedure of diagonalization (see Eq.(4))   we have

\begin{equation}
{{\cal{L}}^D}=-\sum_{i=1}^{3}m_{i}\bar{\nu}_{i}\nu_{i} ,
\label{eq:14 }
\end{equation}

\begin{equation}
\nu_{lL}=\sum_{i=1}^{3}U_{li}\nu_{iL} ,
\label{eq:15 }
\end{equation}

\noindent where $U^{+}U=1 $. It follows from Eq.(14)  that
$\nu_i $ is the field of neutrinos with mass $m_i $. So, if the neutrino
mass term ${\cal{L}}^D $ is present in the Lagrangian, neutrinos are
particles with nonzero
masses and the current fields $\nu_{lL} $ are linear,unitary combinations
of left-handed components of the fields of the massive neutrinos.

It follows from Eq.(13) that in the case under consideration the lepton
numbers $L_e, L_{\mu }, L_{\tau} $
are not conserved separately.However,
 it is easy to see that invariance under
 the global gauge transformation
 $ \nu_{l}~ \rightarrow {e}^{i\alpha} \nu_{l} $,
$ l~ \rightarrow {e}^{i\alpha }l $ holds to the
total lepton number

\begin{displaymath}
L=L_e+L_{\mu }+L_{\tau }
\end{displaymath}

\noindent is conserved, and neutrinos with masses  $ m_i$ are Dirac
particles ($\nu_i $ differs from $\bar{\nu}_i $ by the value of L).

Let us stress in conclusion that in the case of Dirac mass term there is
a full analogy between the quark and lepton sectors of the theory.

\subsection{Majorana mass term.}

\indent
In the Dirac mass term both left-handed and right-handed neutrino fields
enter.Neutrino mass term can be built, however, using left-handed fields
only, if we assume that there are no conserved lepton numbers
$[{\ref{ref.4}}]$.Indeed, let us assume that in the Lagrangian of the system
the following mass term appear

\begin{equation}
{\cal{L}}^M=-\frac{1}{2}\sum_{l',l}\overline{({\nu}_{l'L})^c}
M_{l'l}\nu_{lL}~~+~~h.c.
\label{eq:16}
\end{equation}

\noindent Here $(\nu_{lL})^c=C\bar{\nu}_{lL}^T $ is a right handed component
(C is the matrix of charge conjugation, $C\gamma_{\alpha}^T C^{-1}=
-\gamma_{\alpha}, C^T=-C $) and M is a complex, nondiagonal matrix.From
the Pauli principle it follows that M is a symmetric matrix.For any
symmetric matrix M we have

\begin{equation}
M=(U^+)^T m U^{+}
\label{eq:17}
\end{equation}

\noindent where $U^{+} U=1 $, $m_{ik}=m_{i}\delta_{ik} $,
 $m_{i}>0 $.From Eq.(16) and Eq.(17) it follows that

\begin{equation}
{{\cal{L}}^{M}}=-\frac{1}{2} \sum_{i=3}^{3} m_{i}\bar{\chi}_{i}\chi_{i},
\label{eq:18}
\end{equation}

\noindent where

\begin{equation}
\chi=U^{+}\nu_{L} + (U^{+}\nu_{L})^{c}=
\left( \begin{array}{c}
\chi_{1}\\\chi_{2}\\\chi_{3}
\end{array} \right)
\label{eq:19}
\end{equation}

\noindent and

\begin{displaymath}
\nu_{L}=\left( \begin{array}{c}
\nu_{eL}\\\nu_{{\mu}L}\\\nu_{{\tau}L}
\end{array} \right).
\end{displaymath}

\noindent It is clear
from Eq.(19) that the field $\chi _{i} $ satisfies the condition

\begin{equation}
\chi_{i}^{c}=\chi_{i}
\label{eq:20}
\end{equation}

\noindent which is called Majorana condition.For any fermion field
$\chi(x) $ we have

\begin{eqnarray}
\chi(x) & = & \frac{1}{(2\pi)^{3/2}}\int \frac{1}{\sqrt{2{p}^{0}}}
\Bigl(u^{r}(p) e^{-ipx} c_{r}(p)~+
                                \nonumber \\
   &   &~ +~ u^{r}(-p) e^{ipx}d_{r}^{+}(p)\Bigr) d^{3}p
\label{eq:21}
\end{eqnarray}

\noindent where $c_{r}(p) (~d_{r}^{+}(p)~) $ is the operator of
annihilation of a particle (creation of antiparticle) with
momentum p and helicitiy r.If a fermion field $\chi $ satisfies the
Majorana condition Eq.(20 ), then
we have

\begin{displaymath}
c_{r}(p)=d_{r}(p)
\end{displaymath}

\noindent So the field that satisfies the Majorana condition is a
field of truly neutral (Majorana) particles with spin 1/2
(particle $\equiv $ antiparticle). It is clear from equations Eq.(18)-Eq.(20)
that ${ \cal{L}}^{M} $ is the mass term of  Majorana particles.
This term is not invariant under any global gauge invariance. Futher
from Eq.(19) we will find easily that

\be
\nu_L=U \chi_L~~~
\mbox{or }~~~
\nu_{lL}=\sum_{i=1}^{3} U_{li} \chi_{iL}
\label{eq:22}
\ee

\noindent Thus, if a neutrino mass term is of the form Eq.(16), the current
neutrino fields $\nu_{lL} $ are unitary combinations of the left-handed
components of fields of Majorana neutrinos with definite masses. In this
case there are no conserved lepton numbers.

\subsection{Dirac and Majorana mass term}

\indent
The scheme that corresponds to a Majorana mass term is the most economic
mixing scheme--- only left-handed current neutrino fields enter both the
interaction Lagrangian and the neutrino mass term.The most general
neutrino mixing scheme corresponds to Dirac and Majorana mass term, which
is built with the help of left-handed and right-handed fields under the
assumption that there are no conserved lepton numbers.  Thus, let us
assume that in the Lagrangian of the system the following neutrino
mass term enters $[{\ref{ref.2},\ref{ref.3}}]$

\begin{eqnarray}
{\cal{L}}_{D-M}&=&-~\frac{1}{2}
\sum_{l,l'} \overline{(\nu_{l'L})^c} M_{l'l}^L \nu_{lL}~
-\sum_{l',l} \bar{\nu}_{l'R} M_{l'l}^{D} \nu_{lL}~-
                                  \nonumber\\
               & &-~\frac{1}{2}\sum_{l',l}\bar{\nu}_{l'R} M_{l'l}^{R}
(\nu_{l'R})^c~~+~~h.c.,
\label{eq:23}
\end{eqnarray}

\noindent where $M^L, M^D $ and $M^R $ are 3 $\times $ 3
complex matrixes.

It is clear that, as in the case of Majorana mass term, neutrinos with
definite masses in the case under consideration are Majorana particles.
However, the number of massive particles in this case is twice of the
number of lepton flavours.From Eq.(23), after standard procedure of the
diagonalization of a 6 $\times $6 matrix M we have

\be
{\cal{L}}_{D-M}=-\frac{1}{2} \sum_{i=1}^6 m_i{ \bar{\chi}_i} \chi_i ,
\label{eq:24}
\ee

\noindent where $\chi_i=\chi_i^c $  is the field of Majorana neutrinos
with mass $m_i $. The current fields $\nu_{lL} $ and  fields
$(\nu_{lR})^c=C \bar{\nu}_{lR}^T $~ (left-handed components) are
connected with left-handed components of massive Majorana fields
$\chi_{iL} $ by a unitary transformation

\begin{eqnarray}
\nu_{lL}=\sum_{i=1}^6 U_{li} \chi_{iL}
\nonumber\\
(\nu_{lR})^c=\sum_{i=1}^6 U_{\bar{l}i} \chi_{iL}
\label{eq:25}
\end{eqnarray}

\noindent where U is a unitary 6 $\times $ 6 matrix.

If all masses $m_i $ are small enough, the quanta of the
fields $\nu_{lL} $ are usual flavor left-handed neutrinos and
right-handed antineutrinos,while the quanta of the fields $\nu_{lR} $
are "sterile" right-handed neutrinos and left-handed antineutrinos.
These last particles are sterile in the sense that they do not take
part in the standard weak interaction (fields $\nu_{lR} $ do not enter
the standard Lagrangian of interaction).

The masses $m_i $ and the mixing matrix U are determined by the complex
matrices $ M_L, M_D $  and $ M_R $. One of the most popular mechanism
of neutrino mass generation is based on the assumption that $M_L=0 $ and
elements of $M_D $ are much smaller than the nonzero elements of $M_R $
(see-saw mechanism~ $[{\ref{ref.5}}]$).If the see-saw mechanism is realized,
then the particles with definite masses are three very light Majorana
neutrinos and three very heavy Majorana particles. In the next section
we will consider the see-saw machanism in some detail.

\subsection{See-saw mechanism of neutrino mass generation.}

\indent
 From existing experimental data it follows that the mass of the neutrino
in each generation (if any) is much smaller than the mass of the fermion
in the same generation. The see-saw mechanism of neutrino mass generation
$[{\ref{ref.5}}]$ naturally incorporates this experimental fact. Let
us consider the D-M mass term in the simplest case of one generation.
We have

\begin{eqnarray}
{\cal{L}}^{D-M}& = &
-\frac{1}{2} m_L \overline{(\nu_{L})^c} \nu_L-m_D \bar{\nu}_R \nu_L-
-\frac{1}{2} m_R \bar{\nu}_R (\nu_{R})^c~~+~~h.c.
                  \nonumber\\
               & = &-\frac{1}{2}
\overline{{(\nu_L)^c} \choose {\nu_{R}}}
{}~M~{{\nu_{L}} \choose {\nu_{R}^c}}
{}~~+~~h.c.
\label{eq:26}
\end{eqnarray}

\noindent Here

\be
M= \left( \begin{array}{cc}
m_{L} & m_{D} \\
m_{D} & m_{R}
\end{array} \right),
\label{eq:27}
\ee

\noindent $m_L, m_D, m_R $ are parameters (for simplicity, real). For a
symmetrical matrix M we have

\be
M=O~ m~ O^T,
\label{eq:28}
\ee

\noindent where $O^TO=1 $, $m_{ik}=m_i \delta_{ik} $. From Eq.(26) and
Eq.(28) we have

\be
{\cal{L}}^{D-M}=-\frac{1}{2} \sum_{i=1}^{2} m_i{\bar{\chi}}_i \chi_i ,
\label{eq:29}
\ee

\noindent where

\begin{eqnarray}
\nu_L  =  {\cos{\theta}}\chi_{1L} & + & {\sin{\theta}}\chi_{2L}
\nonumber\\
(\nu_R)^c  =  {-\sin{\theta}}\chi_{1L} & + & {\cos{\theta}}\chi_{2L}.
\label{eq:30}
\end{eqnarray}

\noindent Here $\chi_1 $ and $\chi_2 $ are fields of Majorana neutrinos
 with masses $ m_1, m_2 $. The masses $m_1, m_2 $ and the mixing angle
$\theta $ are connected to the parametres $m_L, m_D $ and $m_R $ by the
relations

\begin{eqnarray}
m_{1,2} & = & \frac{1}{2} {\left|{m_R~ +~m_L~\mp~a} \right|}
\nonumber\\
\sin{2\theta} & = & \frac{2m_D}{a},~~~~\cos{2\theta}=\frac{m_R-m_L}{a}
\label{eq:31}
\end{eqnarray}

\noindent where

\be
a=\sqrt{(m_R-m_L)^2~+~4{m_D^2}}
\label{eq:32}
\ee

Relations Eq.(31) are exact. Let us assume now that

\be
m_L=0,~ m_D\simeq{ m_F},~m_R\gg{ m_F} ,
\label{eq:33}
\ee

\noindent where $m_F $ is the mass of the lepton or quark of
the corresponding generation.From Eq.(31) we have

\be
m_1\simeq\frac{m_F^2}{m_R},~~m_2\simeq{m_R} ,
{}~~\theta\simeq{\frac{m_D}{m_R}}
\label{eq:34}
\ee

\noindent Thus, if the conditions Eq.(33) are satisfied,
the particles
with definite masses are a very light Majorana neutrino with mass
$m_1 \ll m_F $ and a very  heavy Majorana particle with mass
$m_2\simeq m_R $. The current neutrino field $\nu_L $ practically
coincides with $\chi_{1L} $ and $ \chi_2\simeq {\nu_R~+~(\nu_R)^c} $

Usually it is assumed that $m_R=M_{GUT}, M_{GUT} $ is
grand unification scale.
The value of $ M_{GUT} $ depends on the model. Different possibilities
were  considered: from $m_R\simeq {10^{10}} $ GeV
(some intermediate scale)
up to $m_R\simeq {10^{19}} $ GeV (Planck mass). If $m_R $ lies in this
interval, for the heaviest neutrino $\nu_ {\tau} $, for example,
we have

\begin{eqnarray*}
3.10^{-10} \mbox{eV} & \leq{m_{\nu_{\tau}}}~
\leq{3.10^{-1}} \mbox{eV}\\
& \mbox{or} \\
2.10^{-6} \mbox{eV} & \leq{m_{\nu_{\tau}}}~\leq{2.10^3} \mbox{eV}
\end{eqnarray*}

\noindent depending on which see-saw formula we use:
$ m_{\nu_{\tau}} \simeq{\frac{m_{\tau}^2}{m_R}} \mbox{ or }
m_{\nu_{\tau}} \simeq{\frac{m_t^2}{m_R}} $. Let
us stress in conclusion
that the idea of a see-saw mechanism is the following. Assume that in
D-M mass term Dirac masses  are of order of usual fermion masses,
the right-handed Majorana masses, responsible for lepton numbers
violation, are of order of a GUT mass and the left-handed Majorana
masses are equal zero. In such a scheme  neutrinos are Majorana
particles with masses much smaller than masses of the other fermions.
Concrete predictions of neutrino masses depend on the value of the
GUT mass.

\section{Physical consequences of neutrino mixing hypothesis.
Experimental data.}

\subsection{"Direct" method of the measurement of neutrino
mass ($\beta $-spectrum of $ ^3H $).}

\indent
If neutrino masses are different from zero, the hard part of a $\beta $-
spectrum that correspond to emission of soft neutrinos will be
modified. The classical method is the investigation of the spectrum
of the decay

\be
{^3}H~ \rightarrow~ {^3}He~+~e^{-}~+~\bar {\nu}_e
\label{eq:35}
\ee

\nin The electron spectrum in this superallowed decay is determined
by the phase factor.Assuming that the mass of $\nu_e $ is not equal
 to zero, for the electron spectrum in the tritium decay we have

\be
\frac{dN}{dT}=CpE(Q-T)\sqrt{(Q-T)^2~-~m_{\nu}^2}~F(E)
\label{eq:36}
\ee

\noindent Here $ p \mbox{ and } E=m_e+T $ are momentum and energy of
the electron,$ Q $ is the energy release (Q $\simeq $18.6 kev),
$ F(E)$ is the Fermi
function, $m_{\nu} $ is the neutrino mass, $ C $ is a constant.

The experiments on the measurement of neutrino mass by the tritium
method are very complicated. Spectrometer energy resolution,
molecular and other effects must be correctly taken into account.
The measured spectrum is usually fitted with the help of
$ m_{\nu}^2 $  and Q, and additional parametrs that take into
account energy resolution, background and normalization. The data
that have been obtained from recent experiments are presented in
Table 1.

\begin{table}[htbp]
\noindent
Table 1.
Upper bounds for the mass $m_\nu $ obtained
by the tritium method $[{\ref{ref.5}}]$.
\begin{center}
\begin{tabular}{|l|l|}
\hline
Group & Upper bound for $m_\nu$
\\ \hline
Zurich &  $~< 11 eV $
\\
Tokyo &  $~< 13 eV $
\\
LANL &  $~< 9.3 eV $
\\
Mainz &  $~< 7.2 eV $
\\
Livermore & $~< 8 eV $
\\
\hline
\end{tabular}
\end{center}
\end{table}

\nin As it is seen from  Table 1, experiments on the precise
investigation of the hard part of the tritium $ \beta $-spectrum
give only
upper bounds on the mass of neutrino (about 10 $eV $).  Notice that
modern upper bounds of masses of $ \nu_{\mu} \mbox{ and } \nu_{\tau} $
are

\beq
m_{\nu_{\mu}}~<270~~keV
                 \nonumber\\
m_{\nu_{\tau}}~<31~~ MeV \nonumber
\eeq

\subsection{Neutrinoless double $\beta $-decay.}

\indent
The process

\be
(A, Z)~\rightarrow~ (A, Z+2 )~+~e^{-}~+~e^{-}
\label{eq:37}
\ee

\nin is allowed only if the lepton number L is not conserved(M or
D-M mass terms). The neutrino masses enter into the neutrino
propagator.In the case of neutrino mixing

\begin{eqnarray*}
\nu_{eL} =\sum_{i} U_{ei}\chi_{iL}
\end{eqnarray*}

\nin with $\chi_{i} =~C{\bar{\chi}}_i^T $  we have

\beq
\overline{\nu_{eL}(x_1) \nu_{eL}^T}
& = & -\sum{U_{ei}^2} \frac{1+\gamma_5}{2}
\overline{\chi_i(x_1)\chi_i}
(x_2) \frac{1+\gamma_5}{2} C~=
\nonumber\\
& = & \frac{-i}{(2\pi)^4} \sum{U^2_{ei} m_i}
\int{\frac{e^{ip(x_1-x_2)}}{p^2+m_i^2}} dp
\frac{1+\gamma_5}{2} C
\label{eq:38}
\eeq

\nin If the neutrino masses $m_i $ are small enough $ (\leq MeV ),
m_i^2 $ in the integral Eq.(38) can be neglected and all the dependence
of the matrix element on neutrino masses and mixing matrix elements
is in the factor

\be
{}~<m>~=\sum{U_{ei}^2 m_i}
\label{eq:39a}
\ee

\nin Notice that if there is a hierarchy in the lepton sector, similar
to the hierarchy in the quark sector,  the mixing matrix
will be almost diagonal and the main contribution to $ <~m~> $ will
come from the lightest neutrino mass.

More than 30 experiments searching for neutrinoless double $\beta $-
decay ($(\beta\beta)_{0\nu} $-decay) of different nuclei are going on
at present. Up to now there are no positive indications in favour of
the existence of $ (\beta\beta)_{0\nu} $-decay.In  Table 2 some
latest lower bounds on the lifetime of this process are presented.

\begin{table}[htbp]
\noindent
Table 2.
Lower bounds on the lifetime $T_{1/2} \mbox{of}
(\beta\beta)_{0\nu}$ -decay
\begin{center}
\begin{tabular}{|l|l|l|}
\hline
Element & Group & $ T_{1/2}$ \\ \hline
$ ^{76} Ge $ & ITEP/Erevan & $~ >2.\times{10^{24}} y $ \\
$ ^{76} Ge $ & Heidelberg/Moscow &  $~ >.7\times{10^{24}} y $ \\
$ ^{76} Ge $ & UCSB/LBL & $~ > 2.4\times{10^{24}} y $ \\
$ ^{82} Se $ & UCI & $~ > 1.1\times{10^{22}} y $ \\
$ ^{100} Mo $ & LBL/UMN & $~> 1.3\times{10^{22}} y $ \\
$ ^{136} Xe $ & Milano & $~> 2.\times{10^{22}} y $\\
\hline
\end{tabular}
\end{center}
\end{table}

At present new generation of experiments on the search for
$(\beta\beta)_{0\nu} $-decay with enriched $^{76} Ga,~ ^{100} Mo $
and $ ^{136} Xe $ are going on and prepared.
It is expected $[{\ref{ref.7}}]$
that these new experiments will be sensetive
to $~\mid <m>\mid~\simeq $ 0.2-0.3 eV.

\subsection{Neutrino oscillations.}

\indent
If there are neutrino mixing, then neutrino oscillations, that are
analogous to the well known
${K^{0}} \rightleftharpoons {\bar{K^{0}}} $
oscillations,  become possible $[{\ref{ref.8}}]$. Due to the fact that
oscillations are interference phenomena, their search is the most
sensitive method of investigation of neutrino mixing.

We will consider here briefly neutrino oscillations. The vector
of state of flavour neutrino with momentum $\vec{p} $ in the case
of any type of neutrino mixing is given by

\be
\mid{\nu_l>}~=~\sum_i{U^{*}_{li}}\mid i>
\label{eq:39b}
\ee

\nin Here $\mid i> $ is eigenstate of the free Hamiltonian

\be
H_0\mid i>~=~E_i \mid i>,
\label{eq:40}
\ee

\nin where

\be
E_i~=~\sqrt{m_i^2 + p^2 }\simeq{p~+~\frac{m_i^2}{2p}}.
\label{eq:41}
\ee

\nin Thus in the case of mixing, the states of flavour neutrinos
are {\it coherent superpositions } of the states of neutrinos with
definite energies. Notice that this is correct if neutrino mass
differences are small enough.

Assume that at the time t=0 flavour neutrinos $\nu_l $ with momentum
$\vec p $ are produced.At $t>0 $ the state vectors of neutrinos are
given by
\be
\mid{\nu_l}>{_t}~=~\sum\mid\nu_{l'}~>~A_{{\nu_{l'}}{;}{\nu_l}}(t)
\label{eq:43}
\ee

\nin where

\be
A_{{\nu_{l'}}{;}{\nu_l}}(t)=\sum_{i} U_{l'i}e^{-iE_it}U^{*}_{li}
\label{eq:44}
\ee

\nin is the amplitude of transition $\nu_l~\rightarrow~\nu_{l'} $
during the time t. Let us notice that in the case of D-M
mixing in the right-handed side of Eq.(43) the sum over the states of
sterile antineutrinos could enter. So, in the case of neutrino mixing,
at some distance from  the place where $\nu_l $ were produced,
neutrinos $\nu_{l'} $ different from $\nu_l (\nu_{l'}~\neq~\nu_l) $
could be observed. For the probability of the transition
$\nu_l~\rightarrow~\nu_{l'} $ during time $ t\simeq R $ from Eq.(44)
we have

\be
P(\nu_{l} \rightarrow \nu_{l'})= \delta_{ll'}~+~2 Re
\sum_{i<k} U^{*}_{l'i} U_{li} U_{l'k} U_{lk}^{*}
(1-e^{-i\triangle m^{2}_{ik}\frac{R}{p}})
\label{eq:45}
\ee

\nin where $\triangle m_{ik}^{2}=m^2_i~-~m_k^2 $. If
$\triangle m_{ik}^2\frac{R}{p} \ll~1 $ at all $i\neq k $ in this case
$ P(\nu_l \rightarrow \nu_{l'}) \simeq{\delta_{ik}} $.
For neutrino
oscillations to be observed, it is necessary that at least one
neutrino mass squared difference satisfies the condition

\be \triangle m^2~\geq~\frac{p}{R}
\label{eq:46}
\ee

\nin Typical values of the parameter $\frac{p}{R} $ for reactors,
meson factories,
accelerators and the sun are equal respectively to $10^{-2} eV^{2},
10^{-1} eV^{2}, 1 eV^2, 10^{-11} eV^2$.

Experimental data are usually analyzed under the simplest assumptions of
oscillations between two neutrino types
$ \nu_{l}~{\rightleftharpoons}~\nu_{l'}, (l'\neq l) $.
In this case the mixing matrix $U $ has the
form

\begin{displaymath}
U =
\left(\begin{array}{cc}
{}~\cos{\theta} & \sin{\theta} \\
-\sin{\theta} & \cos{\theta}
\end{array} \right)
\end{displaymath}

\nin where $\theta $ is the mixing angle. From Eq.(45) we have

\beq
&& P(\nu_{l} \rightarrow  \nu_{l'}) = \frac{1}{2}\sin^2{2\theta}
(1~-~\cos\frac{\triangle m ^2 R}{2 p})
\nonumber\\
&& P(\nu_{l}\rightarrow \nu_{l}) = P(\nu_{l'} \rightarrow \nu_{l'}) =
1~-~P(\nu_{l} \rightarrow \nu_{l'})
\label{eq:47}
\eeq

\nin There are no indications in favour of neutrino oscillations in
experiments with terrestrial neutrinos. At present new experiments
searching for the transition $\nu_\mu~\rightarrow \nu_\tau $ are going on
$[{\ref{ref.9}}]$ and are planned $[{\ref{ref.10}}]$. We will discuss these
important experiments later on.

\section{Solar neutrinos.}
\subsection{Experimental data. Comparison with the standard solar
model.}

\indent
Experiments on the detecton of neutrinos from the sun are of utmost
importance from the point of view of investigation of the sun as well
as of neutrino properties (neutrino masses and mixing, neutrino magnetic
moment and so on).Detection of neutrinos from the sun began in  1970.
During many years the only solar neutrino experiment was Davis
experiment $[{\ref{ref.11}}]$. Beginning from 1985 solar neutrinos are
detected also by the Kamiokande~ II collaboration $[{\ref{ref.12}}]$.
In  both these experiments only high energy solar
neutrinos, the flux of wich is $\simeq{ 10^{-4}} $
 of the total flux, are detected. A very important step in the
investigation of neutrinos from the sun began at present. First
results of new experiments SAGE $[{\ref{ref.13}}]$ and Gallex$[{\ref{ref.14}}]$
have appeared. In these experiments low energy neutrinos, whose flux
constitute most of the solar neutrino flux, are also detected.

In  Table 3 the main neutrino producing reactions of the solar
$ pp $ and CNO cicles are presented. In the second column of Table 3
we give the neutrino energies and in the third the predicted fluxes
 (in units $10^{10} cm^{-2} sec^{-1} $).

\begin{table}[htbp]
\noindent
Table 3.
Reaction of $pp $ and CNO cicles in wich neutrinos are produced.
\begin{center}
\begin{tabular}{|l|l|l|}
\hline
Reaction & Neutrino energies & Predicted$[{\ref{ref.14}}]$ by the \\
        & (MeV) & SSM flux($10^{10} cm^{-2} sec^{-1} $)

\\ \hline
$pp \rightarrow de^{+}\nu_e $ & 0~-~0.42 & 6.0 \\
$pep \rightarrow d\nu_e $ & 1.44 & $ 1.4\times{10^{-2}} $
\\
$ ^{7}Be~e^{-} \rightarrow {^{7}Li~\nu_e} $ & 0.86(90\%) &
$ 4.9\times{10^{-1}}$
\\
                                        & 0.38(10\%) &  \\
$ ^{8}B \rightarrow {^{8}Be~e^{+}\nu_e} $ & 0~-~14 & $
5.7\times{10^{-4}} $
\\
$ ^{13}N \rightarrow {^{13}C~e^{+}\nu_e} $ & 0~-~1.2 & $
4.9\times{10^{-2}} $
\\
$ ^{15}O \rightarrow {^{15}Ne~e^{+}\nu_e} $ & 0~-~1.73 & $
4.3\times{10^{-2}} $ \\
\hline
\end{tabular}
\end{center}
\end{table}

In the Davis et.al. experiment neutrino are detected by the observation
of $ ^{37}Ar $ production in the reaction

\be
\nu_e~+~^{37}Cl~\rightarrow~e^{-}~+~^{37}Ar
\label{eq:48}
\ee

About 10 atoms of $ ^{37}Ar $ produced during one month are extracted
from 615 t of $C_2Cl_4 $, and in a small proportional counter
K-capture is detected. The average rate of $ ^{37}Ar $ production
in Davis et al. experiment is $[{\ref{ref.11}}]$

\begin{displaymath}
2.10~\pm~0.30~~ SNU
\end{displaymath}

\nin where $1 SNU = 10^{-36} \frac{capture}{atom sec} $. The threshold
of the reaction Eq.(48) is equal to $ E_{th} = 0.814 MeV $.
Thus in the
Davis experiment one detects neutrinos mainly from $^{8}B $-decay
($ \simeq 77\%$) and from $ ^{7}Be~$ K-capture ($\simeq 15\%$).
 From standard solar model it follows that $ ^{37}Ar $
production rate is

\begin{eqnarray*}
8.0 \pm 1.0\quad & SNU & ~~~\mbox{(Bahcall)}~~ [{\ref{ref.14}}] \\
5.8 \pm 1.0 \quad & SNU & ~~~\mbox{(Turck-Chieze)}~~ [{\ref{ref.15}}]
\end{eqnarray*}

\nin Thus, the rate of $^{37}Ar $ production measured in Davis
et al. experiment is
much less than the predicted rate. This inconsistency was called
the solar neutrino problem. In the Kamiokanda II  experiment
$[{\ref{ref.12}}]$ solar neutrinos are detected by the observation
of the process

\begin{eqnarray*}
\nu~+~e~\rightarrow~\nu~+~e
\end{eqnarray*}

\nin The threshold in this experiment is rather high $(E_{th}\simeq
7.5 MeV)$.Thus only $^{8}Be $ neutrinos are detected in K II
experiment. For the ratio of the detected number of events to the
predicted by SSM number it was found

\begin{eqnarray*}
\frac{data}{SSM} & = & 0.46 \pm 0.05 \pm 0.06 ~~~\mbox{(Bahcall)} \\
\frac{data}{SSM} & = & 0.70 \pm 0.08 \pm 0.09 ~~~\mbox{(Turck-Chieze)}
\end{eqnarray*}

\nin Recently the result of two new radiochemocal solar neutrino
experiments GALLEX and SAGE were published $[{\ref{ref.13},\ref{ref.14}}]$.
In these experiments solar neutrinos were detected by the observation
of $^{71}Ge $ production in the reaction

\be
\nu_e~+~^{71}Ga~\rightarrow~e^{-}~+~^{71}Ge
\label{eq:49}
\ee

\nin The threshold of this reaction is  0.233 MeV. Thus, the
$ Ga-Ge $ method allows us to detect solar neutrinos from all sources,
including the main $pp$  source. In the GALLEX  experiment the target
is a water solution of gallium chloride (30.3 tons of $Ga $). The
result published is based on 14 runs of exposition (about 1 year). For
the average rate of $^{71}Ge $ production the value $[{\ref{ref.14}}]$

\begin{displaymath}
( 83 \pm 19 (stat)~ \pm 8\mbox{(syst)} )~~ SNU
\end{displaymath}

\nin was found, which is only about two standard deviations below the
predicted values:

\begin{eqnarray*}
131.5 \pm 7 & SNU & ~~~\mbox{(Bahcall)} \\
124.0 \pm 5 & SNU & ~~~\mbox{(Turck-Chieze)}
\end{eqnarray*}

 From the thermodynamical point of view solar energy is produced by

\begin{displaymath}
4p~\rightarrow~^{4}He+2e^{+}+2\nu_e+27~MeV
\end{displaymath}

\nin So the total flux of solar neutrinos $ I $ is connected to the sun
luminosity L by the relation

\be
I~\simeq \frac{2L}{4\pi R^{2} 27 MeV}
\label{eq:50}
\ee

\nin where R is Sun-Earth distance. For the rate $ Q $ of $^{71}Ge $
production we have

\be
Q~=~\sum_{i}{\bar\sigma}_{i} I_{i}
\label{eq:51}
\ee

\nin where a sum of over-all neutrino sources $i $ is assumed and
${\bar\sigma}_i $ is average cross section of the process Eq.(49).
 From Eq.(50) and Eq.(51) it follows

\begin{displaymath}
Q~\geq {\bar\sigma}_{pp} I \simeq 80\pm 2~~ SNU
\end{displaymath}

\nin
The observed rate satisfies this bound. From our point of view this means
that new experiments are needed to solve the solar neutrino problem if
this problem exists.

In the other $ Ga-Ge $  solar neutrino experiment SAGE metallic
$Ga $ is used
(30 tons of $ Ga $ in the first runs and 57 tons now ). The latest data
of this experiment $[{\ref{ref.13}}]$

\begin{displaymath}
(58 \begin{array}{c}
+17 \\-24 \end{array} \pm 14)~SNU
\end{displaymath}

\nin are in agreement with the GALLEX result.

\subsection{Neutrino mixing and solar neutrinos}

\subsubsection{Introduction}

\indent
In this section we will discuss solar neutrino experiments from the
point of view of the neutrino mixing hypothesis. If there are neutrino
oscillations the beam of  solar neutrinos that initially was a $\nu_e $
beam at the Earth will be described by a mixture of neutrinos of different
types. Radiochemical methods allow us to detect only $\nu_e $. Thus, from
the point of view of neutrino oscillations it is naturally to expect that
the detected $\nu_e $ flux $I_{\nu_e} $ is lower than the initial
$\nu_e $ flux $ I_{\nu_e} $. We have

\be
I_{\nu_e}~=~P(\nu_e~\rightarrow~\nu_e ) I^0_{\nu_e}
\label{eq:52}
\ee

\nin where $ P(\nu_e~\rightarrow~\nu_e) $ is the probability for $\nu_e $
to survive. This last quantity depends on the neutrino mass squared
difference, $ m_{ik}^2=| m_i^2~-~m_k^2 |$ on the elements of the mixing
matrix and on the parameter $\frac{R}{p} $ ($ R $ is the Sun-Earth distance,
$p \simeq E $ is the neutrino energy). If for each $\triangle m^2_{ik} $,
the condition

\begin{displaymath}
\triangle m^2_{ik} \geq \frac{\bar{E}}{R}
\end{displaymath}

\nin is satisfied, from Eq.(45) for the averaged survivial
probability we have

\be
P (\nu_e~\rightarrow~\nu_e)~=~\sum_{i}|U_{ei}|^4
\label{eq:53}
\ee

\nin In the simplest case of the mixing of two neutrino types from
Eq.(53) it follows

\be
P(\nu_e~\rightarrow~\nu_e )~=~\frac{1}{2} (1~+~\cos^{2}{2\theta})
\label{eq:54}
\ee

\nin We see from Eq.(54) that the detected $\nu_e $ flux could be as small
as $\frac{1}{2} $ of the initial flux. This maximal suppression of
$\nu_e $ flux takes place at maximum mixing
$\theta ~\simeq{\frac{\pi}{4}}$.
In the general case of oscillations between n neutrino types we have
from Eq.(53) $[{\ref{ref.2}}]$

\be
P_{min} (\nu_e~\rightarrow~\nu_e)~=~\frac{1}{n}
\label{eq:55}
\ee

\nin This minimum is reached at $|U_{ei}|^2~=~\frac{1}{n} $ ( maximum
mixing ). Thus, in the case of D or M mixing

\begin{displaymath}
P_{min}~=~\frac{1}{3}
\end{displaymath}

\nin and in the case of D-M mixing

\begin{displaymath}
P_{min}~=~\frac{1}{6}
\end{displaymath}

Up to now we have considered only oscillations of solar neutrinos in
vacuum. It was  shown, however, that matter effects could be important.
Now we will discuss these effects.

\subsubsection{Equation of evolution of neutrinos in matter.}

\indent
Let us consider a beam of neutrinos with momentum p and helicity 1.
The vector of state of the beam has the form

\be
|\psi(t) > = \sum_l |\nu_l> a_{\nu_l}(t)
\label{eq:56}
\ee

\nin where $a_{\nu_l}(t) $ is the amplitude of probability to find
$ \nu_l $ at the time t. In vacuum we have

\be
i\frac{\partial{a}(t)}{\partial{t}}~=~H_{0} a(t)
\label{eq:57}
\ee

\nin where

\be
<\nu_{l'}|H_0|\nu_l>~=~\sum <\nu_{l'}|i>E_i<i|\nu_l>
\label{eq:58}
\ee

\nin Here $|i> $ is eigenstate of $H_0 $

\be
H_0|i>~=~E_i|i>
\label{eq:59}
\ee

\nin where

\be
E_i~=~\sqrt{m_i^2~+~{\vec{p}} ^2}~\simeq{p~+~\frac{m_i^2}{2p}}
\label{eq:60}
\ee

\nin Futher, we have

\be
|\nu_l>~=~\sum_{i}|i><i|\nu_l>~=~\sum U_{li}^{*}|i>
\label{eq:61}
\ee

\nin where U is the neutrino mixing matrix in vacuum. From this relation
it follows

\be
<\nu_l|i>~=~ U_{li}
\label{eq:62}
\ee

\nin From Eq.(58)-Eq.(62) for the free Hamiltonian in the flavour
representation we have

\be
H_0~=~UEU^{+}~\simeq{p~+~U\frac{m^2}{2p}U^{+}}
\label{eq:63}
\ee

\nin Now let us discuss the Hamiltonian of interaction of neutrino
with matter. First, let us notice that the neutral current contribution
to the Hamiltonian of interaction is proportional to the unit matrix
($\nu_e-\nu_\mu-\nu_\tau $ symmetry) and can be dropped. The
Hamiltonian of charged current interaction of $\nu_e $'s with electrons
can be written in the form

\be
{\cal{H}}~=~\frac{G}{\sqrt{2}}{\bar{\nu}}_e \gamma_\alpha
(1~+~\gamma_5)\nu_e \bar{e}\gamma^{\alpha}(1~+~\gamma_5)e
\label{eq:64}
\ee

\nin Taking into account that electrons of matter are nonrelativistic
particles for the effective Hamiltonian we have

\beq
(H_I(x))_{\nu_e;\nu_e} & = & \frac{G}{\sqrt{2}}<\vec{p}|{\bar{\nu}}_e
(x)\gamma^{\alpha}(1~+~\gamma_5)\nu_e(x)|\vec{p}>
\nonumber \\
& & <mat|\bar{e}(x)\gamma_\alpha(1~+~\gamma_5)e(x)|mat>=
\nonumber \\
& = & 2\frac{G}{\sqrt{2}}\rho(x)
\label{eq:65}
\eeq

\nin where $ \rho(x) $ is electron density at the point $ x $. Thus,
finally we have the following equation of the evolution of the
neutrino beam in matter $[{\ref{ref.3}}]$

\be
i\frac{\partial{a(x)}}{\partial{x}}~=~(U\frac{m^2}{2p}U^{+}~+~
\sqrt{2}G\rho(x)\beta)a(x)
\label{eq:66}
\ee

\nin Here $x\simeq{t} $ is the distance from the point where neutrinos
where born; $(\beta)_{\nu_e;\nu_e}=1 $, other elements of the
matrix $\beta $ are equal to zero.

\subsubsection{Solutions of the equation of evolution of neutrinos in
matter.}

\indent
In this section we will discuss the solutions of Eq.(66). For
simplicity let us limit ourself to the simplest case of two
neutrino flavours ($ \nu_e $ and, say, $\nu_\mu $ ). For the mixing
matrix in vacuum we have in this case

\be
O~=~
\left( \begin{array}{cc}
{}~\cos{\theta} & \sin{\theta} \\
-\sin{\theta} & \cos{\theta}
\end{array} \right)
\label{eq:67}
\ee

\nin Let us present the Hamiltonian in the form

\begin{displaymath}
H_0+H_I~=~H+\frac{1}{2}Tr(H_0+H_I)
\end{displaymath}

\nin The unit matrix $~\frac{1}{2}Tr(H_0+H_I) $ can be omitted. For
the total Hamiltonian from Eq.(66) and Eq.(67) we have

\be
H(x)~=\frac{1}{4p}
\left( \begin{array}{cc}
2\sqrt{2}G\rho(x)p~-~\triangle m^2\cos{2\theta} & \triangle m^2\sin{2\theta}
\\
\triangle m^2\sin{2\theta} & -2\sqrt{2}G\rho(x)p~+~\triangle m^2\cos{2\theta}
\end{array} \right)
\label{eq:68}
\ee

\nin  where $\triangle m^2=m^2_2-m^2_1.$

Let us transform now the matrix $ H(x) $ to diagonal form. We have

\be
H(x)~=~O(x)E(x)O^T(x)
\label{eq:69}
\ee

\nin where

\begin{displaymath}
O(x)~=
\left( \begin{array}{cc}
{}~\cos\theta(x) & \sin\theta(x) \\
-\sin\theta(x) & \cos\theta(x)
\end{array} \right)
\end{displaymath}

\nin and $ E_{ik}(x)=E_{i}(x)\delta_{ik} $. Here $ E_{1,2}(x) $ are
eigenvalues of $ H(x)$, and $\theta(x) $ is the mixing angle in matter.
 From Eq.(68) and Eq.(69) we have

\beq
\sin{2\theta(x)} & ~=~ & \frac{\triangle m^2\sin{2\theta}}
{\sqrt{X^2+\triangle m^4\sin^2{2\theta}}}
\nonumber \\
\cos{2\theta(x)} & ~=~ & \frac{X}
{\sqrt{X^2+\triangle m^4\sin^2{2\theta}}}
\nonumber \\
 E_{1,2} & = & \mp \frac{1}{4p}\sqrt{X^2+\triangle m^4\sin^2{2\theta}}
\label{eq:70}
\eeq

\nin where

\be
X(x)=\triangle m^2\cos{2\theta}-2\sqrt{2}G\rho(x)E
\label{eq:71}
\ee

As it is seen from Eq.(70) the mixing angle in matrix depends on
$ x $ through
the density $\rho(x) $. In the sun the density is maximal at the center
and decreases in the direction of periphery. Let us assume that at some
point $x=x_R $ we have

\be
\triangle m^2\cos{2\theta}~=~2\sqrt{2}G\rho(x_R)E
\label{eq:72}
\ee

\nin (we will assume that $\triangle m^2>0 $). From Eq.(70) and
Eq.(71) it follows, that at this point the mixing in matter is maximal
$ \theta(x_R)=\frac{\pi}{4} $ for any values of $ \theta $ different
from zero. The condition Eq.(72) is called resonance condition. Let us
notice that at the point $ x=x_R $ the distance between levels is minimal

\be
E_2(x_R)-E_1(x_R)=\frac{\triangle m^2\sin{2\theta}}{2E}
\label{eq:73}
\ee

\nin The resonance condition can be rewritten in the form

\be
\triangle m^2\cos{2\theta}~\simeq 0.7\times{10^{-7}}\frac{\rho_m}
{g/{cm}^3} \frac{E}{MeV} eV^2
\label{eq:74}
\ee

\nin In the center of the sun $\rho_m~\simeq{10^2\frac{g}{cm^3}} $
and for solar neutrinos $ E\simeq MeV $. So for solar neutrinos
the resonance condition is realized at
$ \triangle m~\simeq{10^{-5} eV^2} $.

Let us return now to the evolution equation. From Eq.(66)
and Eq.(69) we have

\be
i \frac{\partial a(x)}{\partial{x}}~=~O(x)E(x)O^{T}(x)a(x)
\label{eq:75}
\ee

\nin Now determine the function

\be
a'(x)~=~ O^{T}(x)a(x)
\label{eq:76}
\ee

\nin For this function we have the following equation

\be
i\frac{\partial a'(x)}{\partial x}~=
(E(x)-iO^{T}(x)\frac{\partial O(x)}{\partial{x}}) a'(x)
\label{eq:77}
\ee

\nin it is easy to show that

\be
-iO^{T}(x)\frac{\partial O(x)}{\partial{x}}~=
\left( \begin{array}{cc}
0 & i \frac{d\theta(x)}{dx} \\
i \frac{d\theta(x)}{dx} & 0
\end{array} \right)
\label{eq:78}
\ee

\nin Further, with the help of Eq.(70) we have

\be
\frac{d\theta}{dx}~=
\frac{\sqrt{2} G\triangle{m^2}\sin{2\theta}E\frac{d\rho}{dx}}
{X^2~+~\triangle{m^4}{\sin^2{2\theta}}}
\label{eq:79}
\ee

\nin So we derivative $\frac{d\theta}{dx} $ is determined by
$\frac{d\rho}{dx} $. Non let us assume that the density varies
slowly enough and

\be
\left| \frac{d\theta(x)}{dx} \right|~\ll~\frac{1}{2}(E_2-E_1)
\label{eq:80}
\ee

\nin In this case we have

\be
i\frac{\partial{a'(x)}}{\partial{x}}~=~E(x)a'(x)
\label{eq:81}
\ee

\nin The solution of this equation has the form

\be
a'(x)~=~e^{-i\int\limits_{x_0}^{x} E_i(x)dx}a_i(x_0)
\label{eq:82}
\ee

\nin The condition Eq.(80) is called adiabatic condition.
If this condition
is satisfied, neutrinos remain at the same energy level.
 From Eq.(76) and
Eq.(82) we obtain the following solution of the evolution equation
in the adiabatic approximation

\be
a(x)~=~O(x) e^{-\int\limits_{x_0}^x E(x)dx}O^{T}a(x_0)
\label{eq:83}
\ee

\nin For the averaged survival probability from Eq.(83) we get

\beq
P(\nu_e~\rightarrow~\nu_e) & = & \sum O^2_{{\nu_e}i}(x)O^2_{{\nu_e}i}(x_0)~=
\nonumber\\
& = & \frac{1}{2}(1~+~\cos{2\theta}(x)\cos{2\theta}(x_0))
\label{eq:84}
\eeq

Let us consider neutrinos that were produced in the region of the sun where
$\rho> \rho(x_R) $. Assume also that $\theta $ is small.
 From Eq.(70) for the
initial mixing angle we have $\theta(x_0)\simeq{\frac{\pi}{2}} $. From
Eq.(84) it follows that survival probability is equal in this case

\be
P(\nu_e~\rightarrow~\nu_e)~\simeq{\frac{1}{2}(1-\cos{2\theta})}~\simeq 0
\label{eq:85}
\ee

\nin Thus , if the adiabatic condition Eq.(80) is satisfied, all the
electron neutrinos that are produced in the region with $\rho>\rho(x_R) $
are transformed into muon neutrinos that cannot be registered by the
radiochemical detectors.

In the general case for the probability of transition $\nu_e \rightarrow
\nu_e $ we have

\be
P(\nu_e~\rightarrow~\nu_e)~=~\sum O^2_{{\nu_e}k}(x)~P_{ki}~O^2_{{\nu_e}i}(x_0)
\label{eq:86}
\ee

\nin where $P_{ik} $ is the probability of transition between levels
with energies $E_i~ \mbox{and}~ E_k, x_0 $ is the initial point and
$ x$ is
the final point. We have from unitarity

\be
P_{12}~=~P_{21},\quad P_{11}~=~P_{22}~=~1-P_{12}
\label{eq:87}
\ee

\nin From Eq.(86) and Eq.(87), for the survival probability we obtain the
following
expression$[{\ref{ref.16}}]$

\be
P(\nu_e~\rightarrow~\nu_e)~=~\frac{1}{2} + \frac{1}{2}\cos{2\theta}(x)
\cos{2\theta}(x_0)(1-2P_{12})
\label{eq:88}
\ee

\nin To determine $P_{12} $ it is necessary to solve the evolution equation.
A rather good approximation is the Landau-Zenner approximation that
is based on the assumption of linear behavior of density near the resonance.
In this approximation$[{\ref{ref.17}}]$

\be
P_{12}~=~e^{-\frac{\pi\gamma(x_R)}{2}}
\label{eq:89}
\ee

\nin where

\beq
\gamma(x_R)~=~\frac{\frac{1}{2}(E_2(x_R)-E_1(x_R))}{{\left|\frac{d\theta}
{dx} \right|}_{x=x_R}} & = &
\nonumber\\
& = & \frac{\triangle m^2}{2E} \frac{\sin^2{2\theta}}{\cos{2\theta}}
\frac{1}{{\left|\frac{d \ln\rho}{dx} \right|}_{x=x_R}}
\label{eq:90}
\eeq

\nin is the adiabaticity parameter. If the adiabatic approximation is
valid, then $ \gamma(x_R)\gg 1 \mbox{ and } P_{12}=0. $ Let
us assume now that $\gamma(x_R)~\ll~1 $ and the resonance condition
is fulfilled. In this case for survival probability from
Eq.(88) we will get

\be
P(\nu_e~\rightarrow~\nu_e)~\simeq \frac{1}{2}(1+\cos{2\theta})~\simeq 1
\label{eq:91}
\ee

\nin Thus,for resonance transition of $\nu_e~\mbox{into}~\nu_\mu $ to
take place, two conditions must be satisfied:\\
1.Resonance condition

\be
\frac{\triangle m^2}{E}~\leq {\frac{2\sqrt{2}G\rho_{0}}{\cos{2\theta}}}
\label{eq:92}
\ee

\nin where $\rho_0 $ is electron density at the center of the sun,

\nin 2.Condition $\gamma(x_R)\geq~1 $

\be
\frac{\triangle m^2}{E}~\geq{\frac{2\cos{\theta}{\left|\frac{d\ln\rho}
{dx} \right|}_{x=x_R}}{\sin^2{2\theta}}}
\label{eq:93}
\ee

\nin  In conclusion we would like
to remark that all existing solar neutrino experimental data could be
described by the MSW mechanism if we assume that the standard solar
model is valid. For the allowed values of the parametrs $\triangle m^2 $
and $\sin^2{\theta} $ the following three regions were
found$[{\ref{ref.18},\ref{ref.19}}]$

\beq
3.2\times{10^{-6}} eV^2 & \leq{\triangle m^2} &
\leq {1.2\times{10^{-5}} eV^2}
\nonumber\\
5.0\times{10^{-3}} & \leq {\sin^2{2\theta }} & \leq 1.6\times{10^{-2}}
\nonumber\\
5.4\times{10^{-6}} eV^2 & \leq {\triangle m^2} &
\leq 1.1\times{10^{-4} eV^2}
\nonumber\\
0.18 & \leq {\sin^2{2\theta}} &  \leq {0.86}
\nonumber\\
10^{-7} eV^2 & \leq {\triangle m^2} & \leq{1.8\times{10^{-6}} eV^2}
\nonumber\\
0.74 & \leq {\sin^2{2\theta}} &  \leq{0.93}
\label{eq:94}
\eeq

\section{Some remarks about future experiments on the search for
neutrino oscillations.}

\indent
As we have discussed before, no indications in favour of neutrino
oscillations were obtained in experiments with terrestrial neutrinos.
The only indications that neutrino masses are different from zero
come from
solar neutrino data. All data available  give some clue in favour
of a possible scenario $[{\ref{ref.20}}]$ that we would like to discuss briefly
in conclusion. Let us assume that there is a mass hierarchy

\be
m_1~\ll {m_2}~\ll {m_3}
\label{eq:95}
\ee

\nin and that $\triangle m^2_{21}=m^2_2~-~m^2_1 $ is so small that

\begin{displaymath}
\frac{\triangle m^2_{21} R}{p}~\ll 1
\end{displaymath}

\nin where R is a terrestrial distance (say $10^{-7} eV^2~<
\triangle m^2_{21}~<10^{-4} eV^2 $). For the average transition
probability  we have

\be
P(\nu_l \rightarrow \nu_{l'})~=~2|U_{l'3}|^2 |U_{l3}|^2,\quad l'\neq l.
\label{eq:96}
\ee

\nin So only the matrix elements that connect lepton flavours
with the heaviestneutrino $\nu_3 $ determine the transition
probabilities in this case.

Now, let us assume also that there is in the lepton sector a hierarchy
of couplings between generations

\be
|U_{e3}|^2~\ll |U_{\mu 3}|^2~\ll |U_{\tau 3}|^2,
\label{eq:97}
\ee

\nin that is analogous  to the hierarchy in the quark sector. From the
unitarity of the mixing matrix we have $|U_{\tau 3}|^2~\simeq 1$. That
means that preferred transitions are those into $\nu_\tau $. Further,
from Eq.(96) and Eq.(97) we have

\beq
P(\nu_e~\rightarrow~\nu_\tau) & \ll & {P(\nu_\mu~\rightarrow~\nu_\tau)}
\nonumber\\
P(\nu_\mu~\rightarrow~\nu_e) & = & \frac{1}{2} P(\nu_e
\rightarrow~\nu_\tau) P(\nu_\mu~\rightarrow~\nu_\tau)~\ll P(\nu_e~
\rightarrow~\nu_\tau)
\label{eq:98}
\eeq

\nin Thus, if there is a hierarchy Eq.(95) and Eq.(97),
$P(\nu_\mu \rightarrow \nu_\tau) $ is the largest transition probability.

New experiments in search for $\nu_\mu \rightarrow \nu_\tau $
oscillations are prepared $[{\ref{ref.8}}]$ and planned $[{\ref{ref.9}}]$
at present. These experiments will be much more sensitive to $\nu_\mu
\rightarrow \nu_\tau $ transitions than the previous one. For example,
if the mass $ m_3~ \mbox{is}~\simeq {1 eV} (\simeq {10 eV}) $ neutrino
oscillations could be seen in these experiments if $\sin^2{2\theta}
\geq  {10^{-2}} (\sin^2{2\theta} \geq {10^{-3}})$.

\vspace{1cm}
{\Large\bf References}

\begin{list}{[\therefs]}{\usecounter{refs}}

\item\label{ref.1}
B. Pontecorvo,
{\it Zh. Eksp. Theor. Fiz.} {\bf 33} (1957) 549;
ibid.{\bf 34} (1958) 247; {\bf 53} (1967) 1717.
\item\label{ref.2}
S.M.Bilenky and B.Pontecorvo,
{\it Phys.Rep.} {\bf 41} (1978) 225.
\item\label{ref.3}
L.Wolfenstein,
{\it Phys.Rev.} {\bf D17} (1978) 2369;
S.P.Mikheyev and A.Yu.Smirnov,
{\it Yad.Fiz.} {\bf 42} (1985) 1441.
\item\label{ref.4}
see review, S.M.Bilenky and S.T.Petcov,
{\it Rev. Mod. Phys.} {\bf 59} (1987) 671.
\item\label{ref.5}
M.Gell-Mann, P.Ramond and R.Slansky,
{\it in Supergravity } ed. by Van Nieuwenhizen and D.Z.Freedman
(North Holland,1979)
T.Yanagida,
{\it Proc. of the Workshop on Unified Theory and Baryon Number
of the Universe}, (Tsukuba, Ikabari,Japan,1979)
\item\label{ref.6}
J.Wilkerson,
Talk given at the {\it "Neutrino"-92} Intern. Conf. Granada, June 8-12,
1992.
\item\label{ref.7}
W.Haxton,
Talk given at the {\it "Neutrino"-92} Intern. Conf. Granada, June 8-12,
1992.
\item\label{ref.8}
B.Pontecorvo,
{\it Zh.Eksp.Theor.Fiz.}{\bf 33} (1957) 549; {\bf 34} (1958) 247
\item\label{ref.9}
N.Armenise et al., CHORUS collaboration,
CERN-SPSC/90-42(1990);
P.Astier et al., NOMAD collaboration,
CERN-SPSC/91-21(1991)
\item\label{ref.10}
K.Kodama et al.
{\it FNAL proposal } {\bf 803} (1991)
\item\label{ref.11}
R.Davis et al.
Intern.Cosmic Ray Conf. January 6-19, Adelaide, Australia,
V.{\bf 7} (1990) 155
\item\label{ref.12}
K.S.Hirata et al.,{\it Phys.Rev.Lett.}{\bf 65} (1990) 1297
\item\label{ref.13}
V.Gavrin, Talk given at 26-th Intern. Conf. on High Energy Physics,
August 5-13,(1992), Dallas, U.S.A.
\item\label{ref.14}
P.Anselmann et al.{\it Phys.Lett.}{\bf 285B} (1992) 376
\item\label{ref.15}
J.N.Bahcall and R.K.Ulrich,
{\it Rev.Mod.Phys.}{\bf 60} (1988) 297.
\item\label{ref.16}
S.Turck-Chieze et al.,{\it Astrophys.J.}{\bf 335}(1988) 425.
\item\label{ref.17}
W.C.Haxton,
{\it Phys.Rev.}{\bf D35} (1987) 2352.
\item\label{ref.18}
P.Anselmann,
{\it Phys.Lett.}{\bf 285} (1992) 390.
\item\label{ref.19}
P.I.Krastev and S.T.Petcov, {\it preprint CERN-TH},6539/92.
\item\label{ref.20}
S.M.Bilenky, M.Fabbrichesi and S.T.Petcov,
{\it Phys.Lett.}{\bf B276} (1992) 223.
\end{list}

\end{document}